\documentclass[prb,aps,amssymb,showpacs,twocolumn]{revtex4-1}
\usepackage{amsmath}
\usepackage{amssymb}
\usepackage{amsthm}
\usepackage{amsfonts}
\usepackage{listings}
\usepackage{enumerate}
\usepackage{latexsym}

\usepackage{psfrag}

\usepackage{bm}
\usepackage{graphicx}
\usepackage{subfigure}

\newcommand{\beq}{\begin{equation}}
\newcommand{\eneq}{\end{equation}}

\input{epsf}

\begin{document}

\tolerance 10000

\newcommand{\vk}{{\bf k}}


\title{Fractional Chern Insulators beyond Laughlin states}

\author{Tianhan Liu$^1$}
\author{C. Repellin$^1$}
\author{B. Andrei Bernevig$^2$}
\author{N. Regnault$^{2,1}$}
\affiliation{$^1$ Laboratoire Pierre Aigrain, ENS and CNRS, 24 rue Lhomond, 75005 Paris, France\\
$^2$ Department of Physics, Princeton University, Princeton, NJ 08544}

\begin{abstract}
We report the first numerical observation of composite fermion (CF) states in fractional Chern insulators (FCI) using exact diagonalization. The ruby lattice Chern insulator model for both fermions and bosons exhibits a clear signature of CF states at filling factors $2/5$ and $3/7$ ($2/3$ and $3/4$ for bosons). The topological properties of these states are studied through several approaches. Quasihole and quasielectron excitations in FCI display similar features as their fractional quantum hall (FQH) counterparts. The entanglement spectrum of FCI groundstates shows an identical fingerprint to its FQH partner. We show that the correspondence between FCI and FQH obeys the emergent symmetry already established, proving the validity of this approach beyond the clustered states. 
We investigate other Chern insulator models and find similar signatures of CF states. However, some of these systems exhibit strong finite size effects.
\end{abstract}

\date{\today}

\pacs{73.43.-f, 71.10.Fd, 03.65.Vf, 03.65.Ud}

\maketitle

\section{Introduction}
Since their recent theoretical proposal\cite{kane-PhysRevLett.95.226801,Bernevig15122006,fu-PhysRevB.76.045302} and their experimental discovery\cite{Koenig02112007,hsieh-nature2008452}, topological insulators have become a major topic in condensed matter physics. The first and simplest example of a topological insulator\cite{haldane-1988PhRvL..61.2015H}, the Chern insulator (CI), is defined by a non-zero Chern number of the occupied bands. It exhibits an integer Hall conductance similar to the integer quantum Hall effect but without an overall magnetic field.

Recently, strong interactions in topological insulators have become a central topic of research. Numerical studies have shown that phases similar to the fractional quantum Hall (FQH) effect can emerge in a flat band CI with a partially filled band and strong interactions, for both fermions \cite{neupert-PhysRevLett.106.236804,sheng-Natcommun.2.389,Bernevig-2012PhysRevB.85.075128,Wu-2012PhysRevB.85.075116,Venderbos-PhysRevLett.108.126405} and bosons\cite{wang-PhysRevLett.107.146803}. These systems have been dubbed Fractional Chern Insulators (FCI).  While the FCI lack the exact magnetic translation symmetries of the FQH\cite{Parameswaran-2011arXiv1106.4025P,goerbig-2012epjb}, equivalent emergent symmetries might arise\cite{Bernevig-2012PhysRevB.85.075128}. Under this assumption, one can build\cite{Bernevig-2012PhysRevB.85.075128} a mapping between the FQH and FCI Brillouin zone which allows for the prediction of the number of topological states and their  quantum numbers in the FCI.

Beside the Laughlin state, Moore-Read (MR) and Read-Rezayi states have also been observed in the FCI using many-body interactions \cite{Bernevig-2012PhysRevB.85.075128,wang-PhysRevLett.108.126805,Wu-2012PhysRevB.85.075116}. Meanwhile, many other FQH states such as the hierarchy or composite fermion (CF) states \cite{Halperin83,Haldane-1983PhysRevLett.51.605,jain89prl199}, have not yet been numerically observed. These states are highly relevant in the context of the FQH effect: they explain (most of) the experimentally observed fractions, including the fractions $\nu=\frac{p}{2p+1}$. It is relevant to study the corresponding series $\nu=\frac{p}{p+1}$ for bosons: in the absence of a promising solid state material candidate for a CI, ultracold atomic gases in optical lattices might pave the way to the experimental implementation of FCI \cite{jo-PhysRevLett.108.045305,Aidelsburger-PhysRevLett.107.255301,Tarruell-2012Natur.483..302T}. The presence of Laughlin-like physics in FCI, analytical arguments\cite{Murthy-2011arXiv1108.5501M} and numerical observation of CF in FQH on a lattice\cite{Moller-2009p184} strongly suggest that such phases should also emerge in FCI. However, as pointed out by two of the authors in Ref.\cite{Wu-2012PhysRevB.85.075116}, not all CI models exhibit a Laughlin-like state, even in the absence of band dispersion or band mixing. In particular the original Haldane's model\cite{haldane-1988PhRvL..61.2015H} with fermions falls into this category. Thus, the existence of hierarchy or CF states in FCI should not be taken for granted.

In this article, we show that some FCI models display a clear signature of a hierarchical or CF states. Our case study focuses on the spin-polarized ruby lattice model~\cite{hu-PhysRevB.84.155116}. Exact diagonalization of this model in the presence of short range interaction gives convincing arguments for the presence of CF states. The nature of these states is probed through their approximated groundstate degeneracy, spectral flow, quasihole and quasielectron excitations, and their entanglement spectrum\cite{li-08prl010504}. The characterization of these properties relies on the FQH-FCI mapping described in Ref.~\cite{Bernevig-2012PhysRevB.85.075128}. We study the stability of these states upon addition of longer range interaction. For all the cases we have studied, adding longer range interaction weakens or even destroys the CF phase. 
 We also discuss the stability of these phases in other CI models. In particular, we find evidence for the existence of CF states in the Kagome lattice model with spin-orbit coupling~\cite{tang-PhysRevLett.106.236802}. The checkerboard lattice model\cite{sun-PhysRevLett.106.236803,sheng-Natcommun.2.389,neupert-PhysRevLett.106.236804,regnault-PhysRevX.1.021014} also shows some signatures of CF states, but with a strong finite size effect. 

 This article is organized as follows. In Sec.~\ref{sec:ModelHamiltonian}, we present the tight-binding Hamiltonian of the ruby lattice model and the effective interacting Hamiltonian that we will numerically study. In Sec.~\ref{sec:NumericalResults}, we present the results of the exact diagonalizations of that model for the groundstate and the excitations of different CF fractions, for both fermions and bosons. We discuss how the low energy spectra of the FCI can be compared to those of the FQH on the torus, thus extending the validity of the existing FQH-FCI mapping beyond model states. In Sec.~\ref{sec:PES}, we use entanglement spectroscopy to probe the topological nature of the groundstate. In Sec.~\ref{sec:LongRangeInt} we analyze the stability of the bosonic CF phase at $\nu = 2/3$ upon addition of longer range interaction.
 
\section{Model Hamiltonian}\label{sec:ModelHamiltonian}
Following Ref.\cite{regnault-PhysRevX.1.021014}, we want to focus on the interaction and the topological properties of the band structure. We remove the effect of band dispersion and band mixing by using the flat band procedure. We start from the original Bloch Hamiltonian $h(\mathbf{k})=\sum_n E_n(\mathbf{k})P_n(\mathbf{k})$ where $E_n(\mathbf{k})$ and $P_n(\mathbf{k})$ are the dispersion  and the projector onto the $n$-th band, respectively. Then we focus on the $i-{\rm th}$ band and consider the effective flat band Hamiltonian $h_{\rm eff}(\mathbf{k})= P_i(\mathbf{k})$. In this picture, we assume that we have infinite band gaps and the $i-1$ first bands are filled and inert.

\begin{figure}[h]
\includegraphics[width=0.55\linewidth]{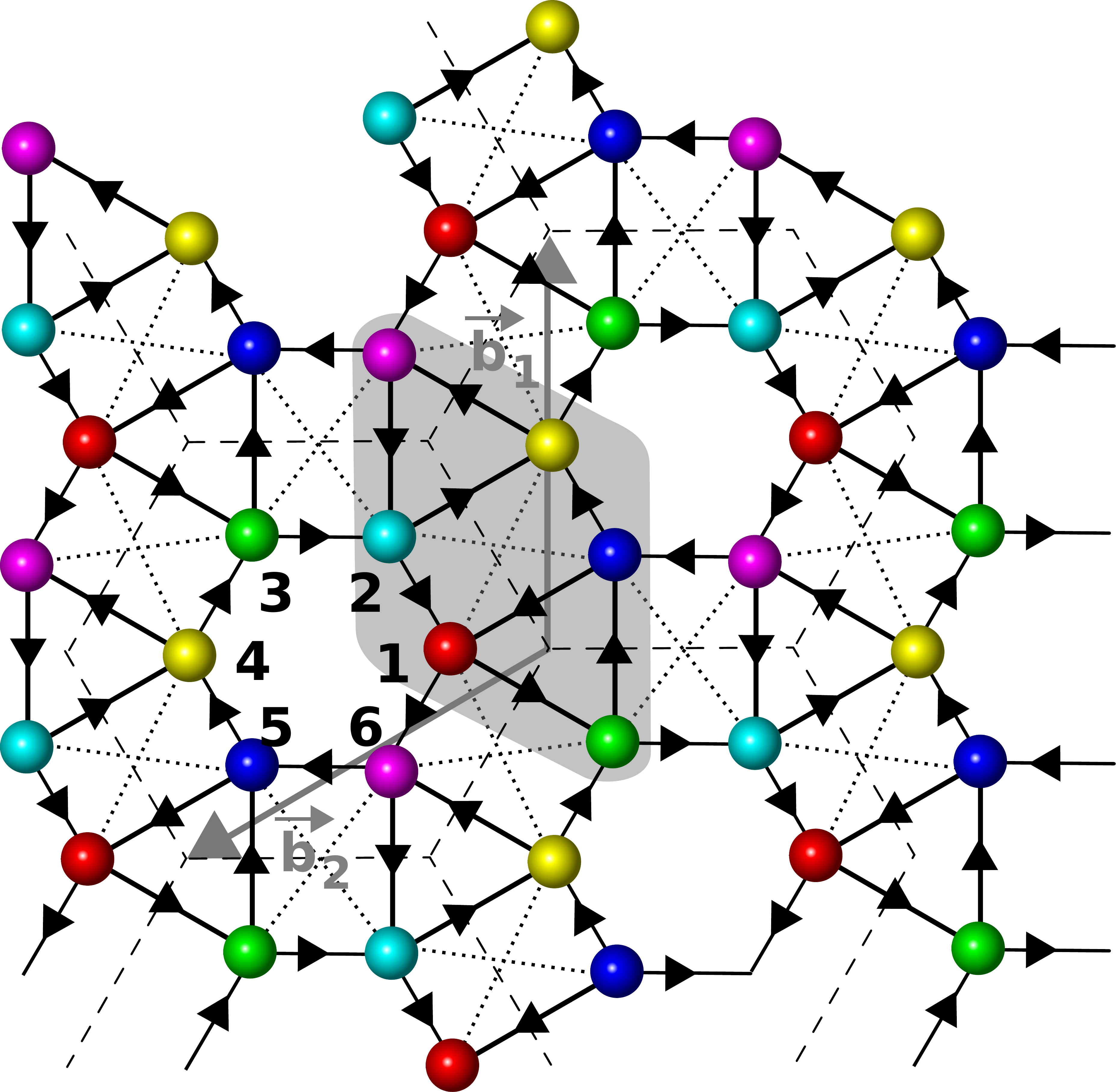}
\caption{The ruby lattice model. The six sublattices are labeled from 1 to 6. The lattice translation vectors are $\boldsymbol  b_1$ , $\boldsymbol  b_2 $. The complex hopping parameters between nearest neighbors are $t = t_r + it_i$ for sites having the same parity and $t_1 = t_{1r} +it_{1i}$ for sites having opposite parity, both in the direction of the arrows. The real hopping parameter on the diagonal of the square is given by $t_4$. } \label{fig:lattice}
\end{figure}

The ruby lattice model has $6$ atoms per unit cell and is spanned by $\boldsymbol b_{1}$ and $\boldsymbol b_{2}$(see Fig. \ref{fig:lattice}). The six sites are denoted from $1$ to $6$. The Bloch Hamiltonian is
\begin{widetext}
\begin{equation}
h(k)=
\begin{bmatrix}
0 &  &  & {\rm h.c.} & \\
t_1^{*} & 0 & & &\\
t & t_1^{*}e^{-i(k_{x}+k_{y})}&0 & & \\
t_{4}(1+e^{ik_{x}}) & t & t_{1}^{*}e^{ik_{x}} & 0 & \\
t^{*} &  t_{4}(1+e^{-i(k_{x}+k_{y})}) & t & t_{1} & 0 & \\
t_1e^{ik_{x}} & t^{*} & t_{4}^{*}(e^{i(k_{x})}+e^{i(k_{x}+k_{y})}) & t &  t_{1}^{*}e^{i(k_{x}+k_{y})}& 0
\end{bmatrix}
\end{equation}
\end{widetext}
where $k_x=\mathbf{k}\cdot \boldsymbol b_1$ and $k_y=\mathbf{k}\cdot \boldsymbol b_2$. The optimal set of parameters for the many body gaps is $t_r=1$, $t_i=1.2$, $t_{1r}=-1.2$, $t_{1i}=2.6$ $t_4=-1.2$ for fermions and $t_r=1$, $t_i=1$, $t_{1r}=-1.4$, $t_{1i}=2.4$ $t_4=-1.46$ for bosons.

 For fermions, we consider nearest neighbor repulsion 
 \begin{equation}
 H^{f}_{\rm int}=U\sum_{<i,j>} : \rho_i \rho_j:
 \end{equation}
 where $::$ denotes the normal ordering. For bosons, we use the two-body Hubbard interaction and neareast neighbor repulsion 
 \begin{equation}
 H^{b}_{\rm int}=U\sum_{i} :\rho_i \rho_i: + V \sum_{<i,j>} : \rho_i \rho_j:
 \end{equation}
 We first restrict ourselves to short range interactions (i.e. $V=0$). In the flat band limit, $U$ is the only energy scale in these systems. We perform exact diagonalizations with $N$ particles on a lattice with $N_x$ (resp. $N_y$) unit cells in $x$ (resp. $y$) direction and periodic boundary conditions. The filling factor is $\nu=N/(N_x N_y)$ with respect to the partially filled lowest energy band of Chern number equal to one.

\section{Numerical results}\label{sec:NumericalResults}

\subsection{Groundstate}\label{subsec:Groundstate}
\begin{figure}[h]
\includegraphics[scale=0.5]{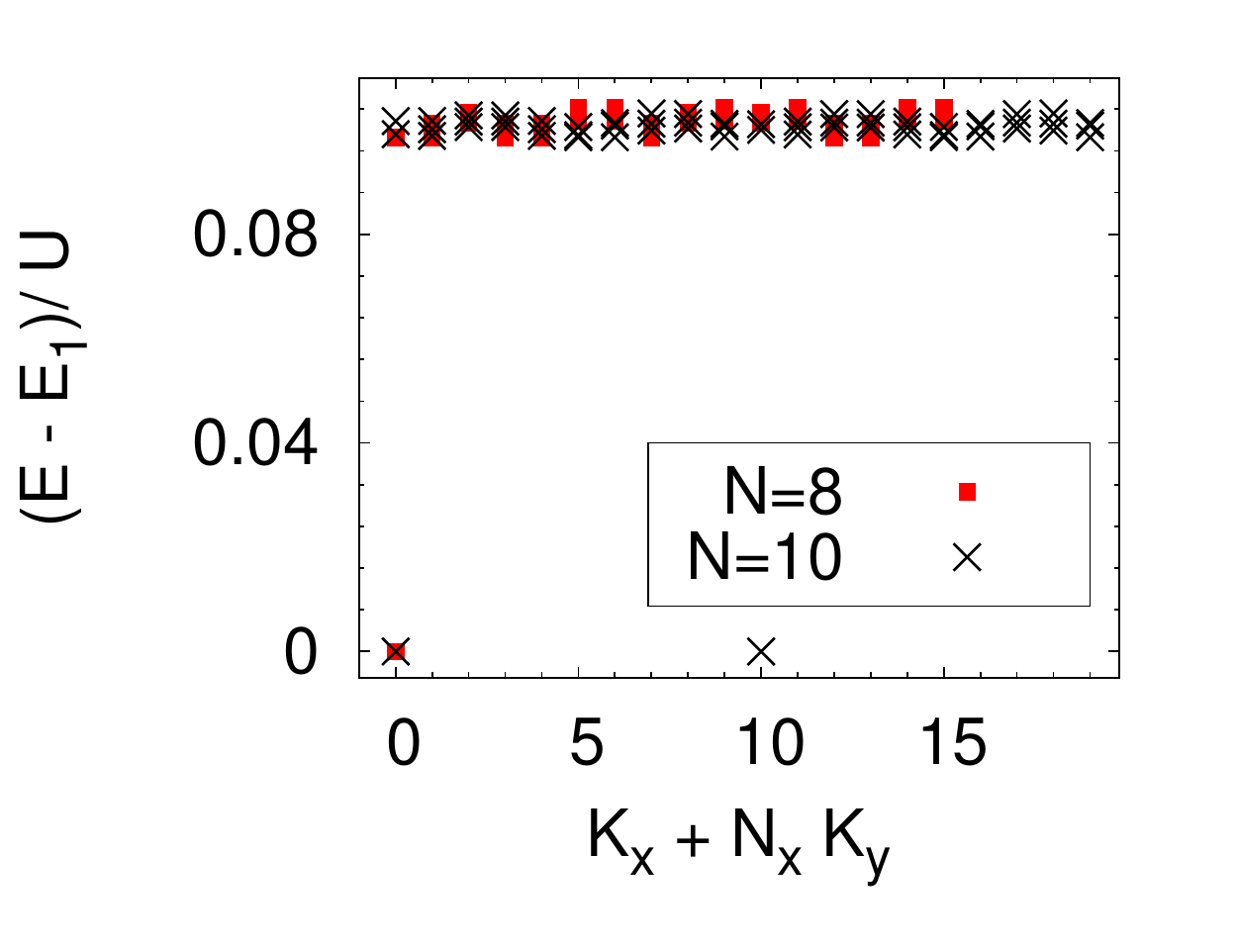}
\caption{Low energy spectrum for $N=8$ and $N=10$ bosons on a ruby lattice with filling factor $\nu=\frac{1}{2}$ on a $N_x=N/2, N_y=4$ lattice. The band parameters have been set to the values given in the article. For $N=8$, the two low energy states occur in the same momentum sector $(0,0)$ and are almost degenerate.}
\label{fig:laughlinruby}
\end{figure}

\begin{figure}[htb]
\includegraphics[scale=0.365]{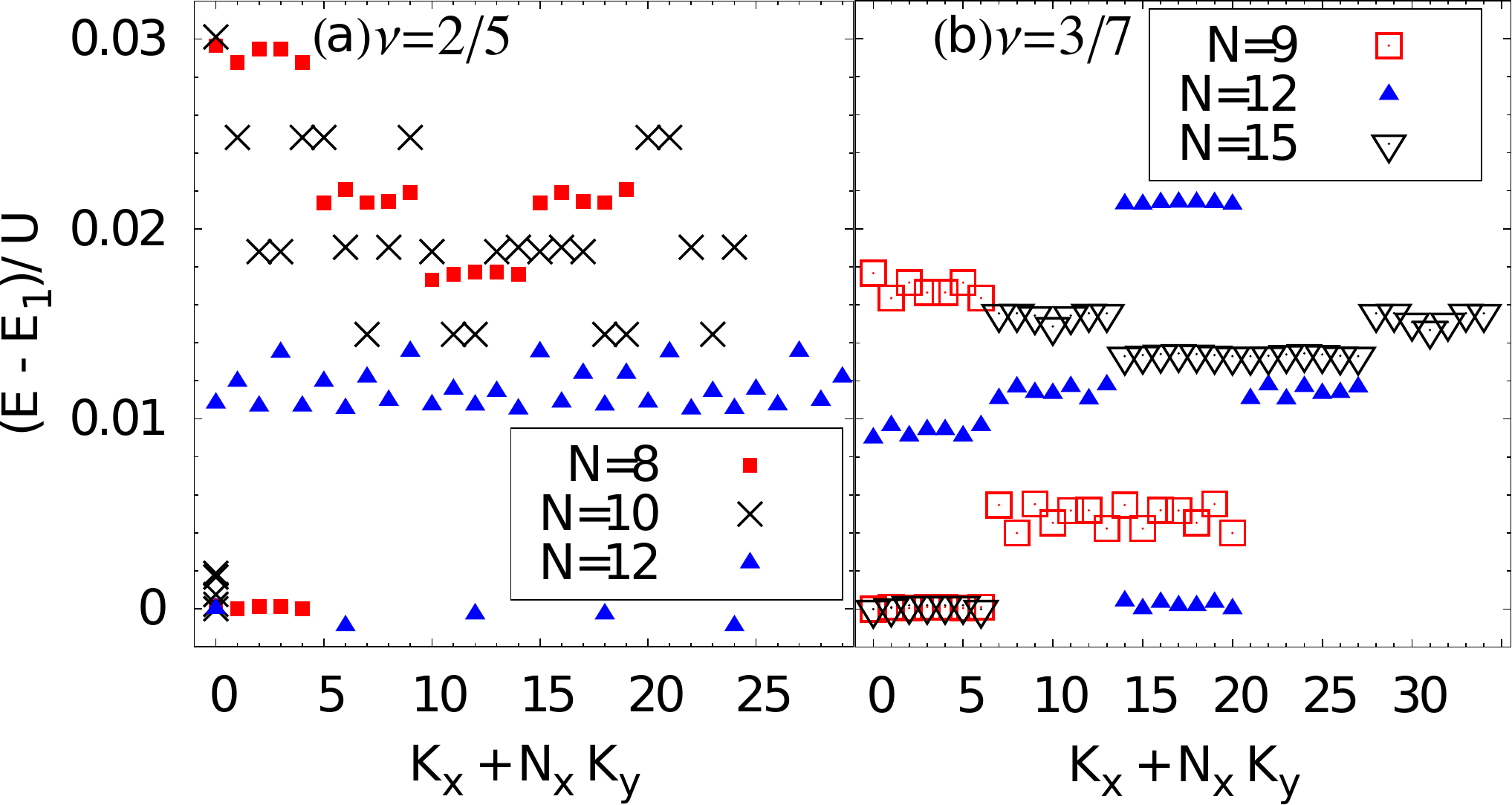}
\includegraphics[scale=0.365]{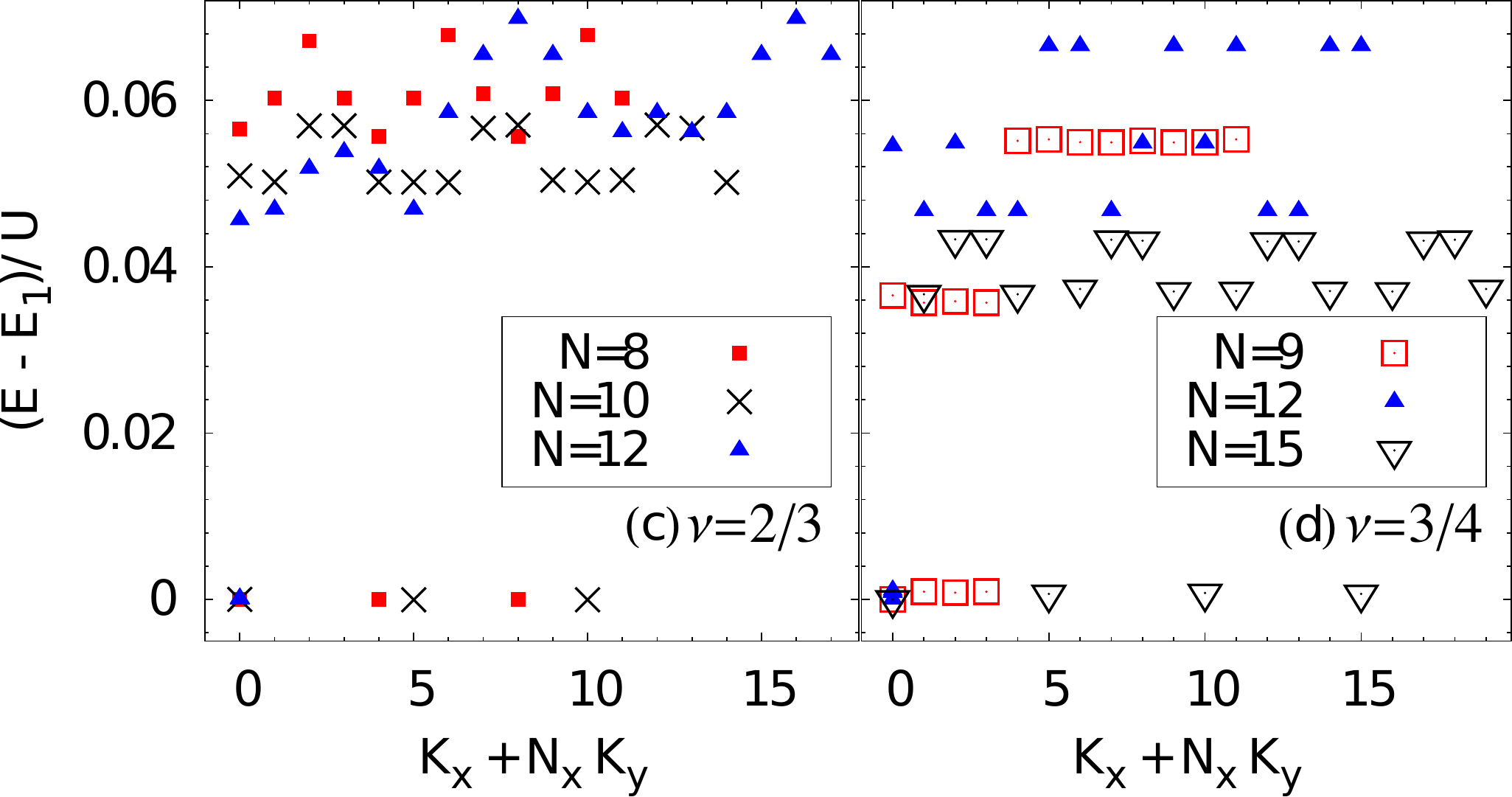}
\caption{Low energy spectra on the ruby lattice for (from left to right, top to bottom) : $N = 8, 10, 12$ fermions at $\nu = \frac{2}{5}$ and $N_x=5, 5, 6$ (a), $N = 9, 12, 15$ fermions at $\nu = \frac{3}{7}$ and $N_x=7$(b), $N = 8, 10, 12$ bosons at $\nu = \frac{2}{3}$  and $N_x= 4, 5, 6$(c) and $N = 9, 12, 15$ bosons at $\nu = \frac{3}{4}$ and $N_x=4, 4, 5$(d). The energies are shifted by $E_1$, the lowest energy for each system size. We only show the lowest energy per momentum sectors in addition to the degenerate groundstate (except for $N=15$ where we only show one energy state per momentum sector). For $N=12$ in Figs. (c) and (d), all groundstates occur in the momentum sector $K_x=K_y=0$ and are almost perfectly degenerate.}
\label{ground_state_ruby}
\end{figure}

We first focus on the groundstate at filling factor $p/(np+1)$ with $n=1$ for bosons and $n=2$ for fermions. For the FQH effect on the torus geometry, these states have a $(np+1)$-fold degeneracy. The absence of exact magnetic translation symmetries in FCI will lift this degeneracy\cite{Parameswaran-2011arXiv1106.4025P,goerbig-2012epjb,Bernevig-2012PhysRevB.85.075128} but we expect to observe a low energy manifold separated by a gap from higher energy excitations. In Ref.~\cite{Wu-2012PhysRevB.85.075116}, only the fermionic case at $\nu=1/3$ has been considered. For bosons on such system at filling $\nu=1/2$, there is also a strong Laughlin-like phase as can be observed in Fig.~\ref{fig:laughlinruby}. The energy splitting between the almost twofold degenerate groundstate is of the order of $10^{-5}$. 
Fig. \ref{ground_state_ruby} shows the results on the ruby lattice for fermions at $\nu=2/5$ (Fig. \ref{ground_state_ruby}a), $\nu=3/7$ (Fig. \ref{ground_state_ruby}b), and for bosons at $\nu=2/3$ (Fig. \ref{ground_state_ruby}c), $\nu=3/4$ (Fig. \ref{ground_state_ruby}d). We observe a nearly $(2n+1)$-fold degeneracy for $N=8$, $10$, $12$ particles (see Fig. \ref{ground_state_ruby}) at $2/(2n+1)$,  and $3n+1$-fold degeneracy for $N=9$, $12$, $15$ particles at $3/(3n+1)$. 

\begin{figure}[h]
\includegraphics[width=0.98\linewidth]{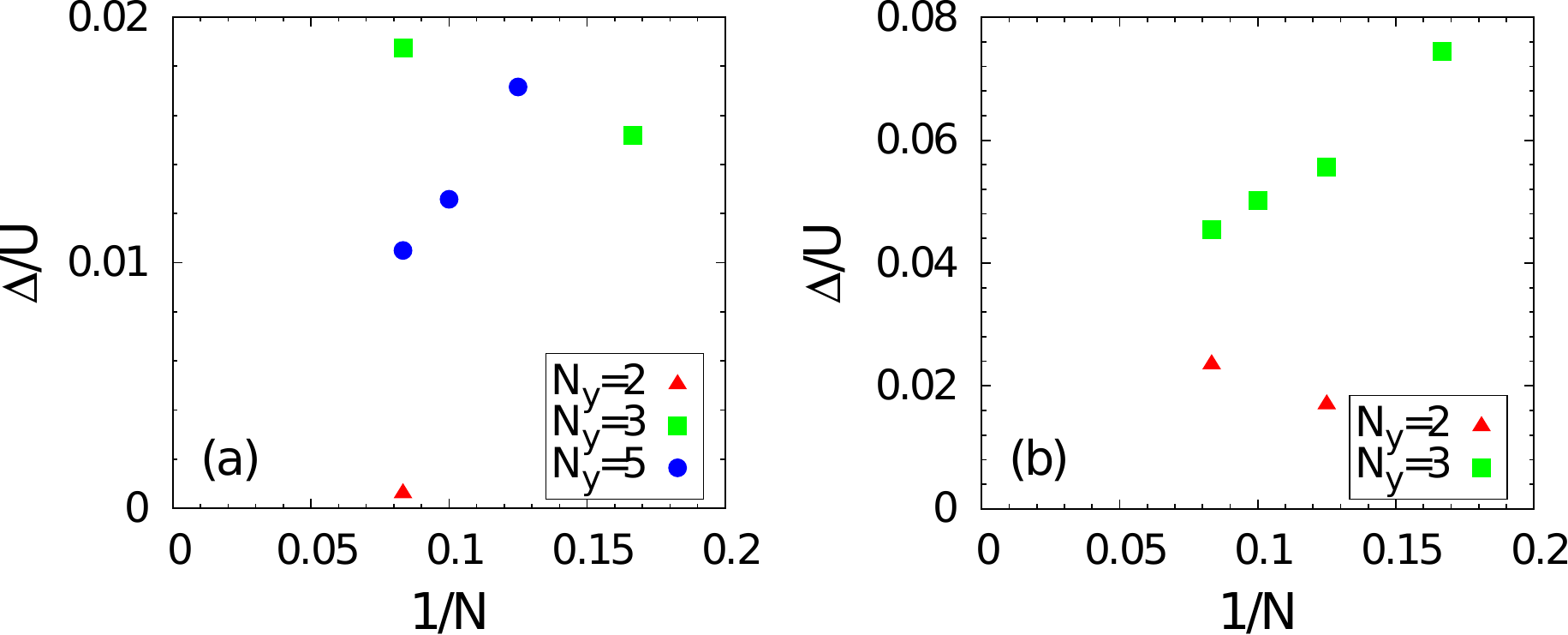}\\
\includegraphics[width=0.98\linewidth]{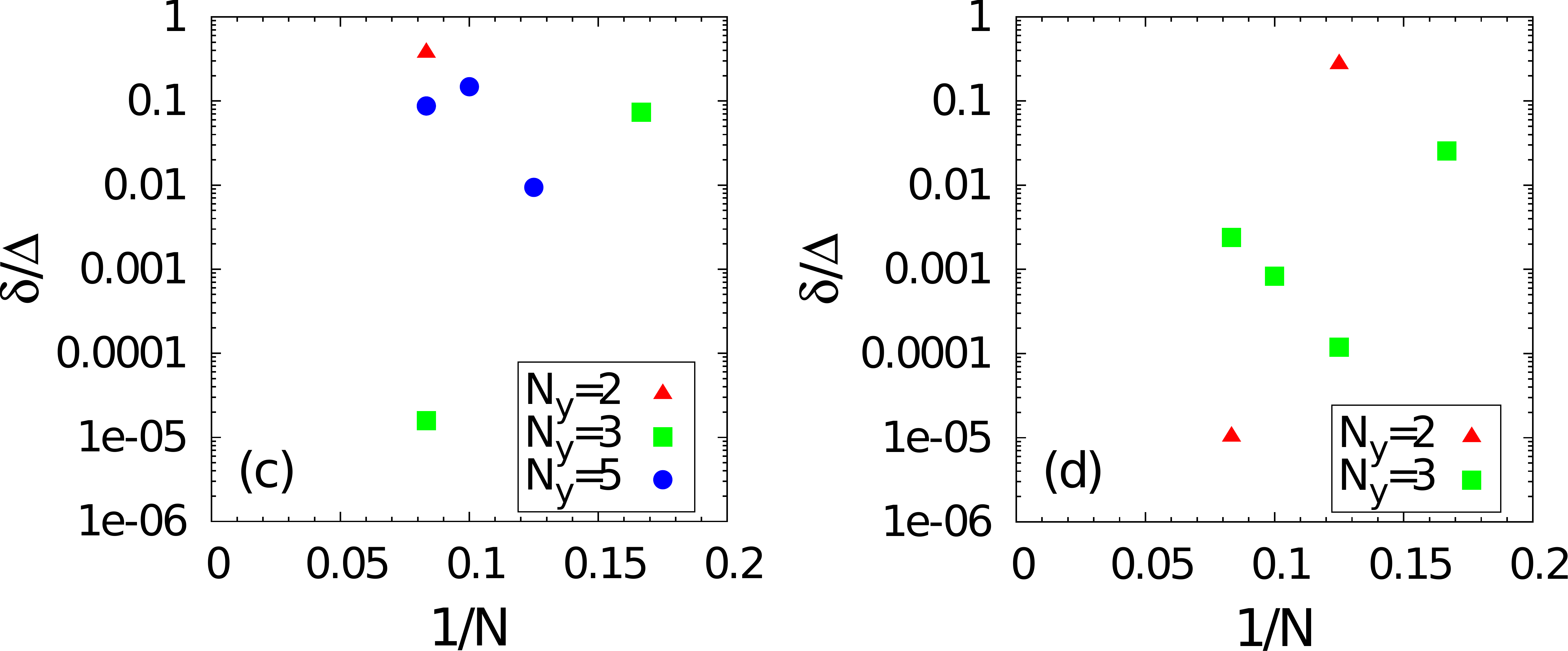}
\caption{(a) Gap $\Delta$ as a function of $1/N$ for fermions on the ruby lattice at filling factor $\nu=2/5$ with $N_x=(5N)/(2N_y)$. (b) Similar plot for bosons at $\nu=2/3$ with $N_x=(3N)/(2N_y)$. (c) Energy spread $\delta E_1$ as a function of $1/N$ for fermions on the ruby lattice at filling factor $\nu=2/5$ with $N_x=(5N)/(2N_y)$. (d) Similar plot for bosons at $\nu=2/3$ with $N_x=(3N)/(2N_y)$.} \label{fig:gapspreadruby}
\end{figure}

As already pointed out in Ref.~\cite{regnault-PhysRevX.1.021014}, the gap may actually increase with the system size if the torus aspect ratio is close to unity (as observed in Fig. \ref{ground_state_ruby}b between $N=9$ and $N=12$). Such a dependence prevents attempts to perform a gap extrapolation. Nevertheless, we show in Figs.~\ref{fig:gapspreadruby}a and \ref{fig:gapspreadruby}b the gap $\Delta$ as a function of the number of particles for fermions (resp. bosons) at filling factor $\nu=2/5$ (resp. $\nu=2/3$). Another important quantity to characterize the groundstate manifold is its energy spread $\delta E_1$, defined as the difference between the largest and the smallest energy in its spectrum.  Indeed, we can distinguish a low energy manifold from the higher energy states  when the condition $\delta E_1 < \Delta$ is satisfied. Note that for the FQH case, $\delta E_1$ is strictly equal to zero. The numerical results are displayed in Figs.~\ref{fig:gapspreadruby}c and \ref{fig:gapspreadruby}d.

We have performed a similar study for several other models, starting with the Kagome lattice model~\cite{tang-PhysRevLett.106.236802} for fermions at $\nu=2/5$ and bosons at $\nu=2/3$. In both cases, the groundstate degeneracy lifting is more important than the one observed for the ruby lattice model. 
The situtation is even more dramatic for the checkerboard lattice model\cite{sun-PhysRevLett.106.236803,sheng-Natcommun.2.389,neupert-PhysRevLett.106.236804,regnault-PhysRevX.1.021014}. Even if one tunes the band structure parameters, the gap exhibits strong finite size effects. These additional numerical results are available in the supplemental material \cite{SuppMeta}.

The almost degenerate groundstates manifold appears in well defined momentum sectors which can be deduced from the FQH to FCI mapping\footnote{In the case of the groundstate, an approach similar to the one developed in Ref.~\cite{regnault-PhysRevX.1.021014} with root partition, works also for CF states. In that case, one should consider the thin torus limit} developed in Ref.\cite{Bernevig-2012PhysRevB.85.075128}. We denote ${\cal N}^{\rm FQH}\left(k_x , k_y \right)$ the degeneracy of the FQH groundstate or quasihole manifold for $N$ particles and $N_\Phi = N_x \times N_y$ flux quanta in the $\left(k_x , k_y \right)$ momentum sector. Similarly, we call ${\cal N}^{\rm FCI}\left(k_x , k_y \right)$ the  approximate degeneracy for the corresponding FCI case. Then the FQH to FCI mapping is given by
\begin{eqnarray}
{\cal N}^{\rm FCI}\left(k_x , k_y \right)=\sum_{k_x', k_y'=0}^{N-1}&& \frac{N_{x0} N_{y0}}{N_0} {\cal N}^{\rm FQH}\left(k_x',k_y'\right) \label{folding}\\
	& & \times \delta_{k_x' {\rm mod} N_{x0}, k_x} \delta_{k_y' {\rm mod} N_{y0}, k_y}, \nonumber
\end{eqnarray}
where $N_{x0} = {\rm GCD}(N, N_x)$, $N_{y0} = {\rm GCD}(N, N_y)$ and  $N_0 = {\rm GCD}(N, N_\Phi)$, ${\rm GCD}$ denotes greatest common divisor. Eq.~\ref{folding} works for the quasihole states as well.

Until now, this mapping has only been used for states such as the Laughlin or Moore-Read ones, where the counting per momentum sector of the FQH wavefunctions for both the groundstate and the quasihole excitations can be derived from a generalized Pauli principle\cite{Haldane-PhysRevLett.67.937}. In the case of CF states, such a principle is missing. Nevertheless, we can perform a direct comparison between the energy spectrum of a FQH system on a torus and its counterpart in a FCI assuming Eq.~\ref{folding} still holds. There is no known exact Hamiltonian for the CF states, but we can rely on realistic interactions, such as the Coulomb interaction for fermions or the delta interaction for bosons. In both cases, the CF states correctly describe the low energy spectrum.

\subsection{Quasiholes and quasielectron excitations}\label{subsec:QHAndQE}
Probing the topological nature of these states requires going beyond the observation of the $nq+1$ quasi-degeneracy of the groundstate manifold : a charge density wave (CDW) in a finite size system may exhibit a similar feature.  One can perform an adiabatic flux insertion and check that these $nq+1$ states flow into each other. We have observed this property for all the cases previously mentioned. Although this does not discard the possibility of a CDW, this is additional argument in favor of a fractional state. The hallmark of the FQH is the unique nature of its excitations which are realized by inserting or removing flux quanta. In the case of FCI, bulk excitations are nucleated by adding or removing unit cells.

\begin{figure}[h]
  \includegraphics[scale=0.35]{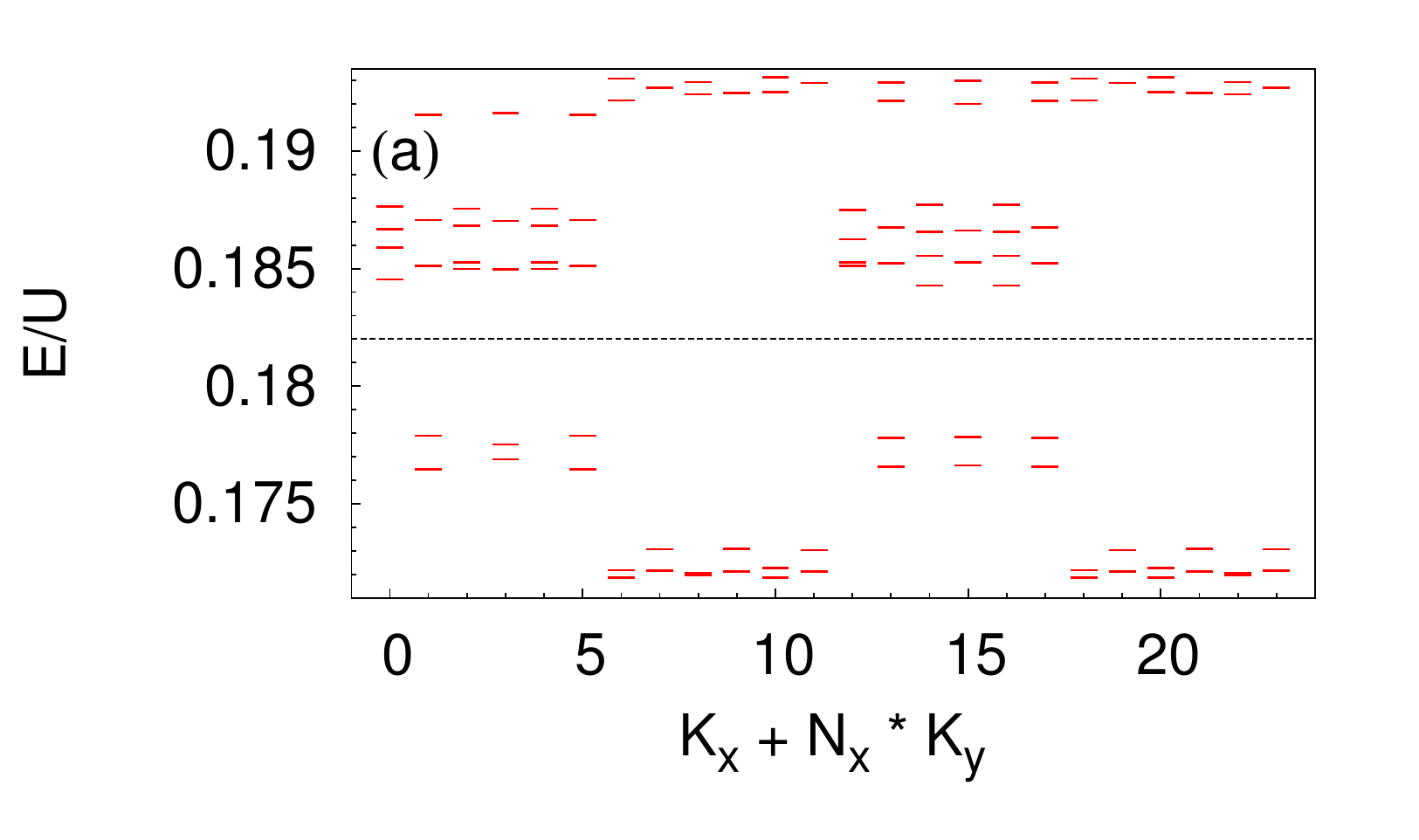}
  \includegraphics[scale=0.35]{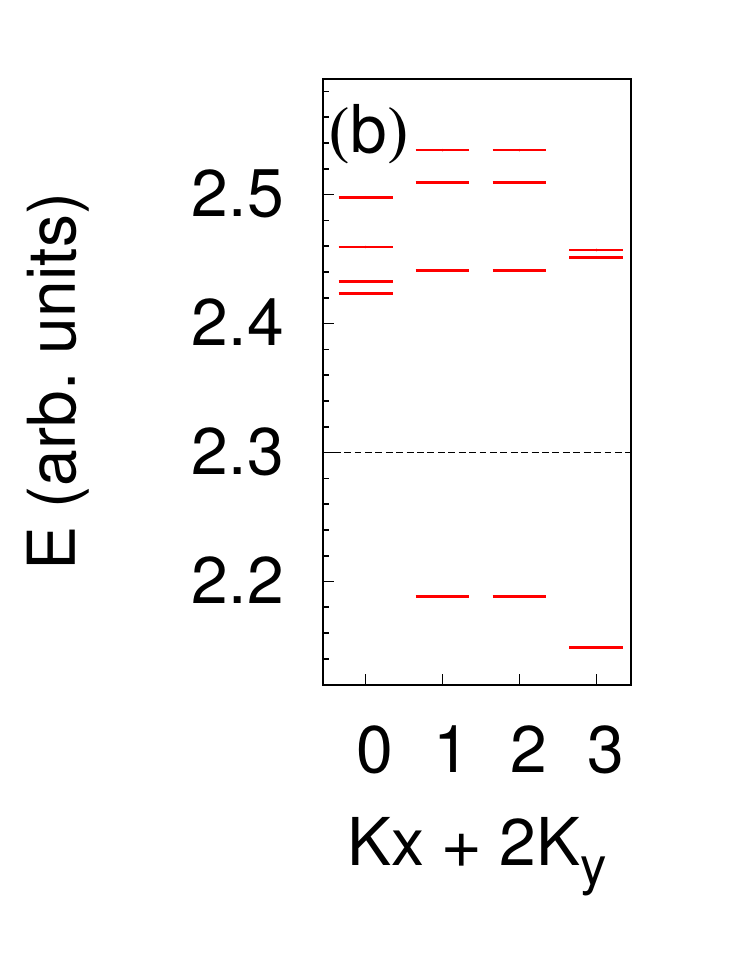}
 \caption{Low energy spectra for $N=10$ fermions on a $(N_x,N_y)=(6,4)$ ruby lattice (a) and its FQH counterpart for fermions on a torus with the Coulomb interaction and $N_{\Phi}=24$ flux quanta (b). Both systems are one flux (or one site) less than the $\nu=\frac{2}{5}$ groundstate and thus embed two quasielectrons. For the FQH plot, we only display the sectors that are not related by the 12-fold center of mass translation symmetry. The number of states per momentum sector below the gap (depicted by the dashed line) in the FCI spectrum can be deduced from the FQH spectrum using the FQH-FCI mapping.}\label{fermion_quasielectron}
\end{figure}

For example, one can generate two quasielectrons at filling $\nu=2/5$ with $N=10$ fermions by looking at the system on a $6*4$ ruby lattice (i.e. one unit cell less than the groundstate). Fig. \ref{fermion_quasielectron}a shows the corresponding low energy spectrum which possesses a clear gap. We compare it to the energy spectrum of the identical FQH system (see Fig. \ref{fermion_quasielectron}b): the FQH-FCI mapping perfectly relates the two countings below the gap. We have observed similar results for both bosons and fermions, on the ruby and Kagome models, for both quasielectrons and quasiholes. While the observation of these features is required, it is not a definitive proof of the topological nature of these states. As discussed in Ref.\cite{Bernevig-2012arXiv1204.5682B}, a CDW state would have a similar counting structure.

\section{Particle entanglement spectrum}\label{sec:PES}
\begin{figure}[h]
\includegraphics[scale=0.39]{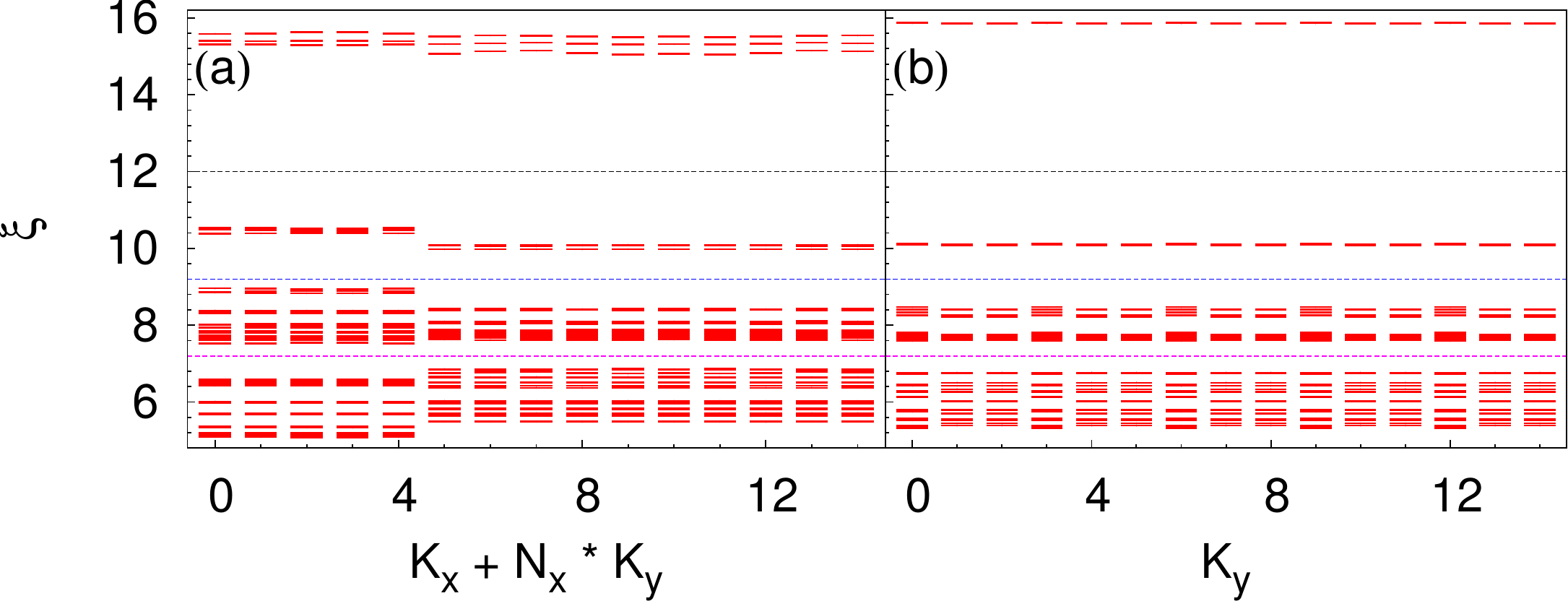}
 \caption{PES of bosonic FCI on the ruby lattice (a) and FQH system with delta interaction (b)  at  $\nu=\frac{2}{3}$, $N=10$ and $N_A=3$. We observe several gaps in both spectra (each depicted by a line). The counting below each of these gaps obeys the FQH-to-FCI mapping. The counting below the largest (and topmost) gap corresponds to the number of Moore-Read quasihole states. The number of states below the middle gap is related to the Gaffnian state.}
\label{PES}
\end{figure}

The nature of the groundstate can be probed through the entanglement spectrum\cite{li-08prl010504} (ES). This technique allows to unveil the information encoded in the groundstate wavefunction. In particular, it has been shown\cite{Bernevig-2012arXiv1204.5682B} that the ES differentiates between a CDW and a Laughlin state. We use the particle entanglement spectrum\cite{sterdyniak-PhysRevLett.106.100405} (PES). For a $d$-fold degenerate state $\{|\psi_i>\}$, we consider the density matrix $\rho=\frac{1}{d}\sum_{i=1}^{d}|\psi_i><\psi_i|$. We divide the $N$ particles into two groups $A$ and $B$ with respectively $N_A$ and $N_B$ particles. Tracing out on the particles that belong to $B$, we compute the reduced density matrix $\rho_A={\rm Tr}_B \rho$. This operation preserves the geometrical symmetries of the original state, so we can label the eigenvalues $\exp(-\xi)$ of $\rho_A$ by their corresponding momenta. A typical PES is shown in Fig. \ref{PES}a, where the $\xi$'s (
generally called energies) are plotted as a function of the momentum.  For FQH model states, the number of non zero eigenvalues in $\rho_A$ matches the number of quasihole states for $N_A$ particles and the same number of flux quanta as the original state. The quasihole counting is characteristic of each topological state. Thus the PES acts as a fingerprint of the phase.

In FCI, one expects to observe a low energy structure similar to the one of the model state with a gap to higher energy excitations. Such a feature has been shown for Laughlin and MR-like states in FCI \cite{regnault-PhysRevX.1.021014, Wu-2012PhysRevB.85.075116}. There is no known formula to count the number of quasihole states for a given CF state: we directly compare the PES of the FCI and the FQH. In the FQH case, the PES does not always exhibit a clear entanglement gap. Nevertheless, the case $N_A=3$ is quite insightful : we observe several gaps (see Fig. \ref{PES}b for the bosonic case). The counting below the most prominent one corresponds to the counting of the MR quasihole states. This is a consequence of the vanishing property of both the CF states at $\nu=2/(2n+1)$ and the MR state\cite{sterdyniak-1367-2630-13-10-105001}. A second gap is present and isolates a low energy structure associated with the Gaffnian state\cite{yoshioka-PhysRevB.38.3636,simon-07prb075317}. The FCI PES also displays a very 
similar structure related to the one of the FQH using the mapping described previously. This is a strong indication that the two systems describe a similar phase.

\section{Effect of longer range interaction}\label{sec:LongRangeInt}
\begin{figure}
\includegraphics[scale=0.39]{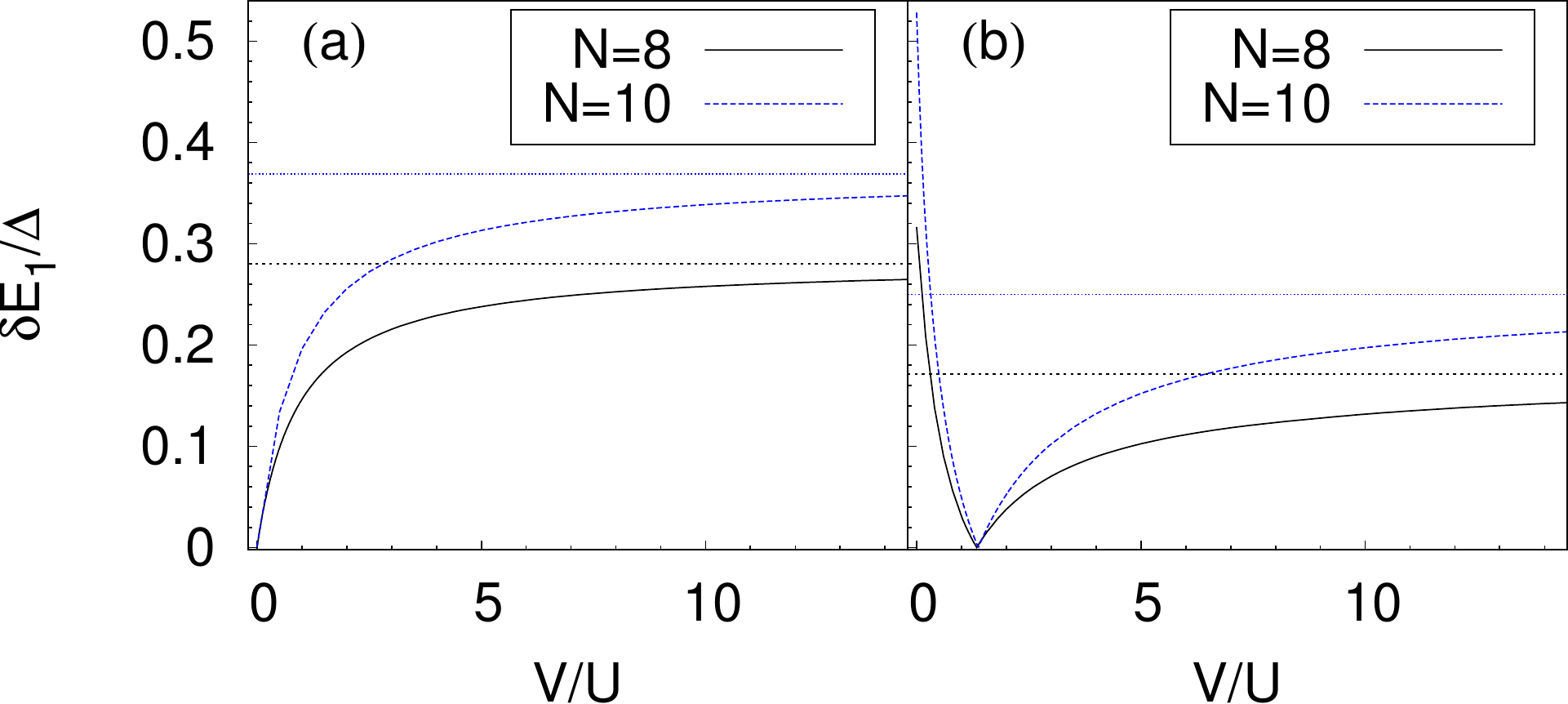}
\caption{Evolution of $\frac{\delta E_1}{\Delta}$ with increasing two-body nearest neighbor repulsion $V$ in the bosonic ruby lattice system for $N=8$ and $N=10$ at filling factor $\nu=2/3$ on a $N_x=N/4, N_y=4$ lattice. (a) We use the band parameters defined in this paper that produce the lowest ratio $\frac{\delta E_1}{\Delta}$ when only the on-site interaction is applied. (b) We have set the band parameters to the values described in Ref.~\cite{hu-PhysRevB.84.155116} that lead to the flattest lowest band in the ruby lattice model. For both Figs a and b, the horizontal line is guide for the eyes that gives the asymptotic value when $U=0$.}
\label{fig:Vinteraction}
\end{figure}

We study the stability of the bosonic CF state, focusing on the $\nu=\frac{2}{3}$ case with an additional nearest-neighbor repulsion ($V>0$). The bosonic CF state at $\nu = \frac{2}{3}$ is characterized by a nearly 3-fold degenerate groundstate separated from the rest of the spectrum by a manybody gap $\Delta$, which can either be direct or indirect. Upon introduction of a nearest-neighbor repulsion, the gap $\Delta$ decreases, and the groundstate degeneracy is lifted even more, with an energy spread $\delta E_1$ defined as the difference between the largest and the smallest energy in its spectrum. We have looked at the ratio $\frac{\delta E_1}{\Delta}$ when increasing the nearest neighbor interaction scale $V$. As shown in Fig.~\ref{fig:Vinteraction}a, increasing $V$ yields to a progressive destruction of the CF state. While we can still distinguish between the threefold degenerate groundstate and the first excited state at large $V$, the asymptotic value of $\frac{\delta E_
1}{\Delta}$ seems to become larger with the number of particles. As a result, CF-like states are not stable towards longer range interaction. This is in  agreement with a similar observation at the filling factor $\nu=1/3$\cite{sheng-Natcommun.2.389,Wu-2012PhysRevB.85.075116}. This result is not really surprising. Even in the FQH regime, the addition of longer range interaction to the Laughlin model interaction (using the Haldane's pseudo-potentials~\cite{Haldane-1983PhysRevLett.51.605}) will destroy the Laughlin phase. While there is no pseudo-potential description in FCI, the instability that we observe in Fig.~\ref{fig:Vinteraction} has the same origin than the one in FQH. Note that using another set of band parameters (those identified in Ref.~\cite{hu-PhysRevB.84.155116} to produce a flattest lowest band in the ruby lattice model), we find a slightly different result. Adding some amount of $V$ (here  $V\simeq 1.35 U$) seems to minimize $\frac{\delta E_1}{\Delta}$ (see Fig.~\ref{fig:Vinteraction}b). This 
minimum is of the same order as in the case of the optimal band parameters. This suggests that one can play with a combination of either the interaction form or the band structure to optimize a given FCI. 

\section{Conclusion}
We have studied the FCI at filling factors $p/(2p+1)$ for fermions and $p/(p+1)$ for bosons, focusing on the $p=2$ and $3$ cases. By conjecturing that the FQH-FCI mapping\cite{Bernevig-2012PhysRevB.85.075128} is valid for any low level energy or entanglement energy structures, we have provided convincing numerical evidence that CF like states emerge in several lattice models for both fermions and bosons. 
Similarly to what is observed in the Laughlin physics, long range interactions tend to weaken these phases.
Surprisingly, strong finite size effects can arise from models displaying a very clear Laughlin phase. Further research should focus on understanding what key ingredients of a lattice model give rise to such physics  and generalizing the CF states to FCI with a higher Chern number\cite{wang-PhysRevB.84.241103,Barkeshli-PhysRevB.84.205134,Wang-2012arXiv1204.1697W,Trescher-2012arXiv1205.2245T,Yang-2012arXiv1205.5792Y}.

\emph{Acknowledgements}
NR thanks G. M\"oller, A. Sterdyniak, Z. Papic and J. Jain for useful discussions. BAB was supported by Princeton Startup Funds, NSF CAREER DMR-095242, ONR - N00014-11-1-0635, Darpa - N66001-11-1-4110, Darpa - 62459-PH-DRP, Packard Foundation and Keck grant. NR was supported by  NSF CAREER DMR-095242, ONR - N00014-11-1-0635, Packard Foundation and Keck grant.

\bibliography{fcicf}

\newpage

\begin{center}
{\bf Supplementary Material to ``Fractional Chern Insulators beyond Laughlin states''}
\end{center}

\begin{figure}[htb]
\includegraphics[width=0.98\linewidth]{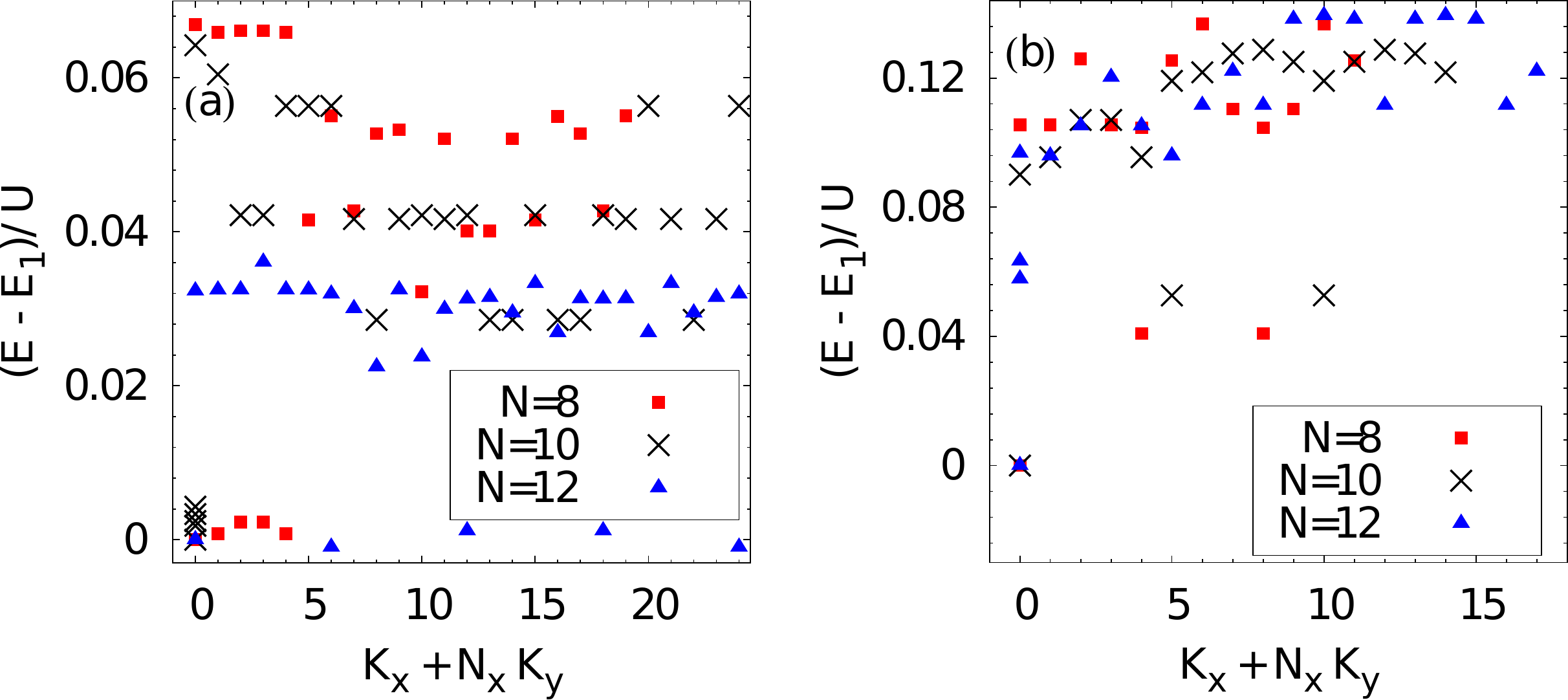}
\caption{Low energy spectra on the Kagome lattice for $N = 8, \ 10, \ 12$ fermions at filling factor $\nu = \frac{2}{5}$ and $N_x=5, 5, 6$ (a) and $N = 8, \ 10, \ 12$ bosons at filling factor $\nu = \frac{2}{3}$ and $N_x= 4, 5, 6$ (b). The energies are shifted by the lowest energy $E_1$ for each system size. We only show the lowest energy per momentum sectors in addition to the approximate degenerate groundstate manifold.}
\label{ground_state_kagome}
\end{figure}

\begin{figure}[htb]
\includegraphics[scale=0.34]{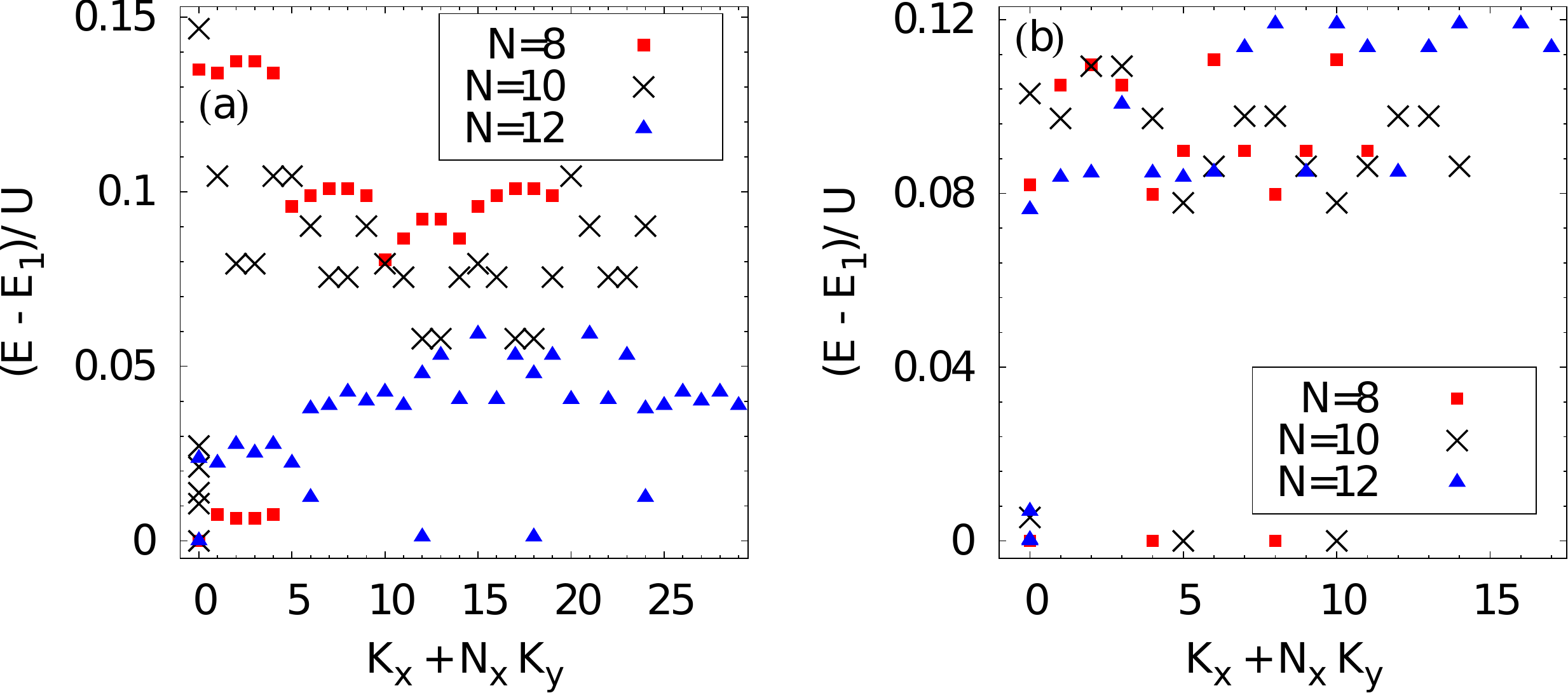}
\caption{(a) Low energy spectrum of $N = 8, \ 10, \ 12$ fermions on a checkerboard lattice at filling factor $\nu = \frac{2}{5}$. (b) Low energy spectrum of $N = 8, \ 10, \ 12$ bosons on a checkerboard lattice at filling factor $\nu = \frac{2}{3}$. We use the same conventions as in Fig.~\ref{ground_state_kagome}.}
\label{fig:checkerboardenergy}
\end{figure}

\begin{figure}
\includegraphics[scale = 0.4]{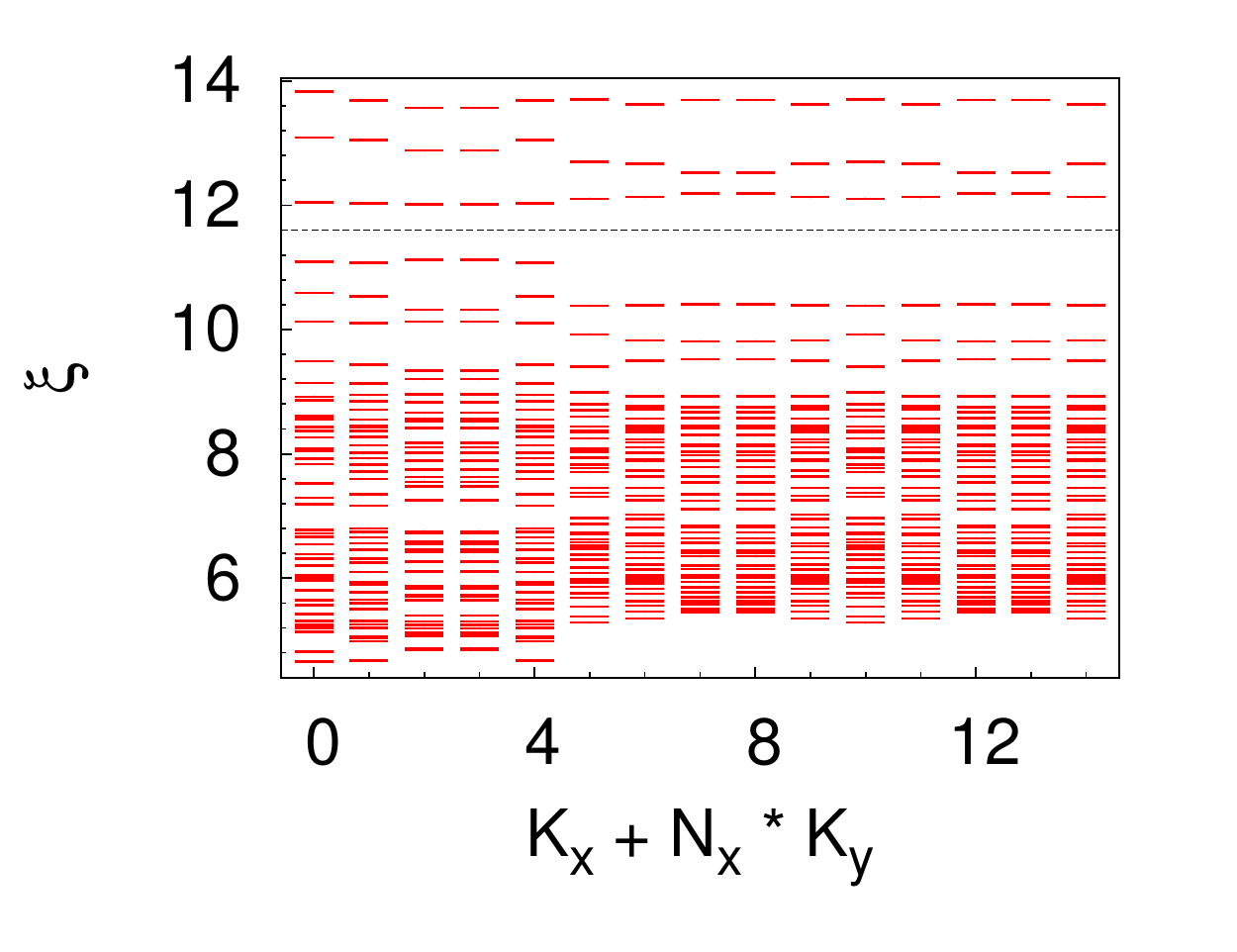}
\caption{PES for $N=10$ bosons on a $N_x=5, N_y=3$ checkerboard lattice, with $N_A = 3$. The entanglement spectrum does not show any clear gapped structure except for the one (materialized by the black line) related to the Moore-Read counting.}
\label{fig:checkerboardPES}
\end{figure}

In this Supplementary Material, we provide additional numerical results that might be relevant to a more specialized audience. In addition to the ruby lattice model, we have also studied the Kagome~\cite{tang-PhysRevLett.106.236802} and checkerboard~\cite{sun-PhysRevLett.106.236803,sheng-Natcommun.2.389,neupert-PhysRevLett.106.236804,regnault-PhysRevX.1.021014} lattice models for both fermions and bosons at filling factors $\nu=p/(np+1)$. Fig.~\ref{ground_state_kagome} shows the energy spectrum for $\nu=2/5$ (fermions) and $\nu=2/3$ (bosons) on the Kagome lattice. We observe a low lying manifold separated from higher energy states by a manybody gap. The $np+1$  dimension of the ground state manifold is characteristic of a CF-like state. The ground state degeneracy lifting is more important than in the ruby lattice case, especially for bosons. Fig.~\ref{fig:checkerboardenergy} shows the energy spectrum for $\nu=2/5$ (fermions) and $\nu=2/3$ (bosons). For the fermionic case, the approximate degeneracy quickly 
deteriorates when increasing the system size despite tuning the band structure parameters. The bosonic case seems to have a better behavior but investigating the eigenstates with the PES reveals a less favorable picture. As shown in Fig.~\ref{fig:checkerboardPES} for $N=10$ particles, the PES does not display any gapped structure, thereby preventing the identification of the groundstate as a topological FQH state.

\begin{figure}[h]
\includegraphics[scale=0.47]{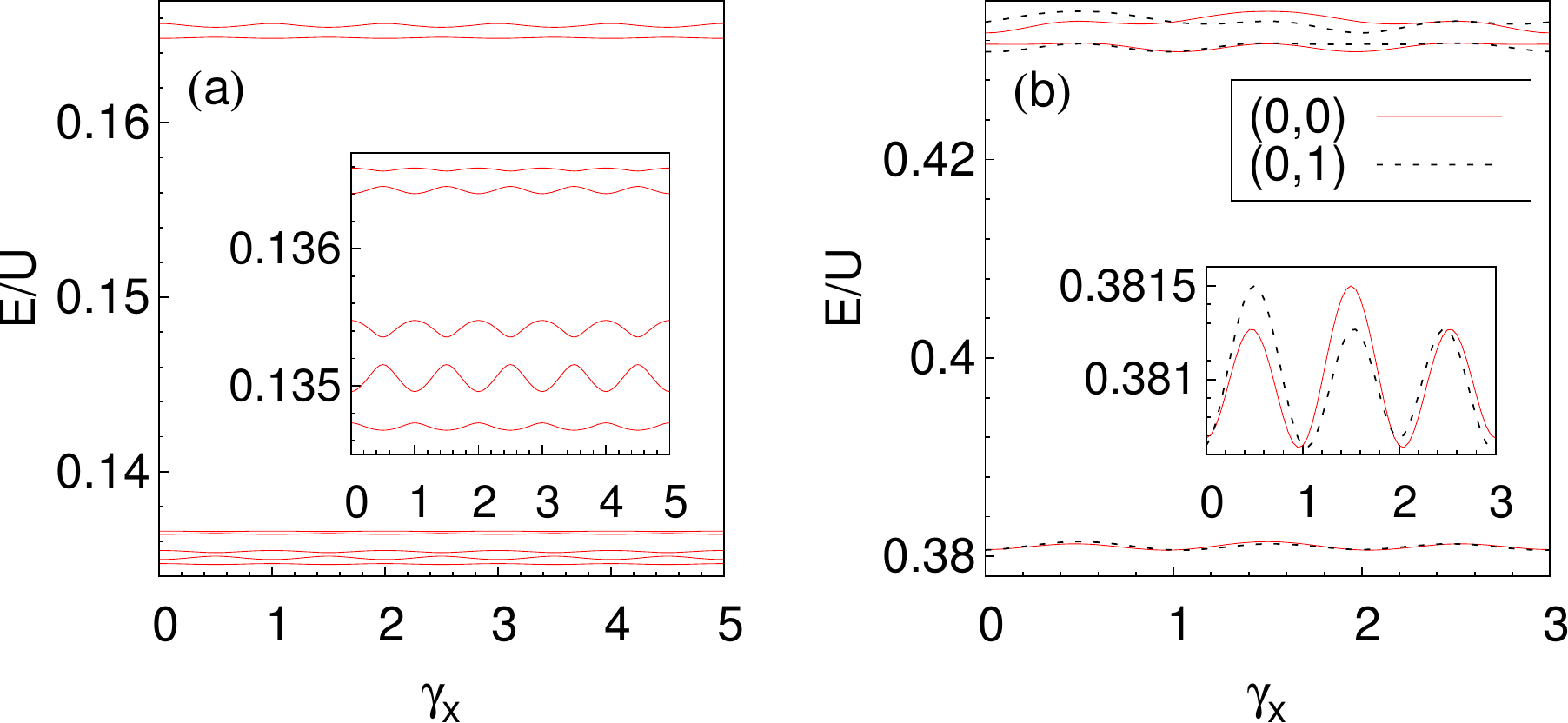}
\caption{(a) Evolution of the low-lying states of the ruby lattice model in momentum sector $(0,0)$ with $N=10$ fermions on a $(N_x,N_y)=(5,5)$ lattice at $\nu=\frac{2}{5}$ upon flux insertion along the $x$ direction. $\gamma_x$ counts the number of inserted flux quanta. We only show the momentum sector $(K_x,K_y) = (0,0)$ where the almost 5 fold degenerate groundstate lies. (b) Evolution of the low-lying states of the ruby lattice model in momentum sectors $(0,0)$ and $(0,1)$ for $N=10$ bosons on a $(N_x, N_y) = (5,3)$ lattice ($\nu=\frac{2}{3}$) upon flux insertion along the $x$ direction. The third groundstate in the $(0,2)$ sector is not shown since it is related to the $(0,1)$ sector by the inversion symmetry. In both Figs. a and b, the inset is a zoom on the low energy part of the spectrum.} \label{spectralflow}
\end{figure}

\begin{figure}[h]
  \includegraphics[scale=0.35]{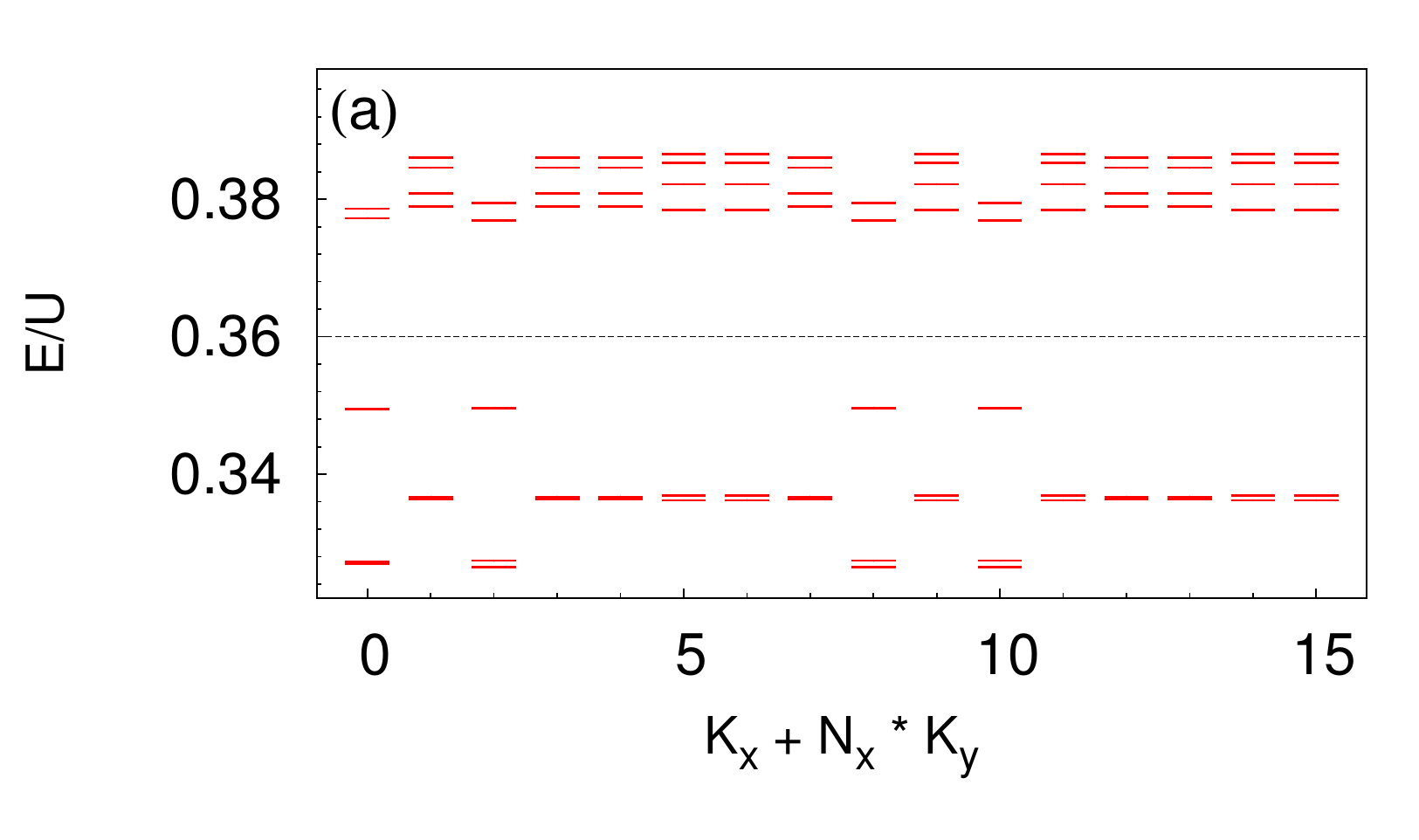}
  \includegraphics[scale=0.35]{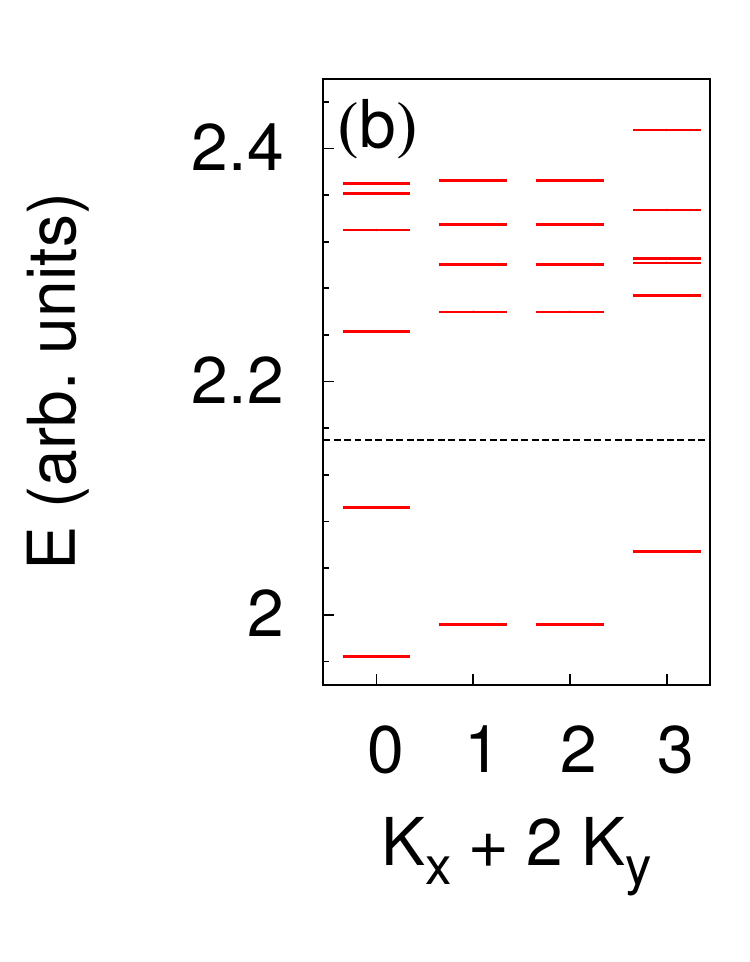}
 \caption{Low energy spectra for the $N=10$ bosons on the $N_x=N_y=4$ ruby lattice (a) and the FQH equivalent for bosons on the torus with delta interaction and $N_{\Phi}=16$ flux quanta (b). Both systems have one more flux (or one more site) than the $\nu=\frac{2}{3}$ groundstate and thus embed two quasiholes. For the FQH plot, the Brillouin zone is partially represented; we only display the sectors that are not related by the 8-fold center of mass translation symmetry. The number of states per momentum sector below the gap (materialized by the dashed line) in the FCI spectrum can be deduced from the FQH spectrum using the FQH-FCI mapping.}\label{boson_excitation}
\end{figure}

We now give more details about the FCI on the ruby lattice.
 As mentioned in the article, we have checked that upon flux insertion the $\nu = p/(np+1)$ groundstate manifold does not mix with higher energy states. Also, the insertion of $np+1$ fluxes restores the original configuration. This can be observed for $N=10$ particles in Fig.~\ref{spectralflow}a for fermions at $\nu=2/5$ and in  Fig.~\ref{spectralflow}b for bosons at $\nu=2/3$. We have also looked at the excitation energy spectrum both in the case of quasielectrons and quasiholes. Similarly to the case discussed in the article, one can generate two quasiholes at filling $\nu = 2/3$ with $N=10$ bosons by looking at the system on a $4*4$ ruby lattice (see Fig.~\ref{boson_excitation}a). Once again we compare the low energy structure of this system to the FQH one (see Fig.~\ref{boson_excitation}b) and check that the number of states per momentum sector is predicted by the FQH-FCI mapping.

\begin{figure}[h]
\includegraphics[scale=0.39]{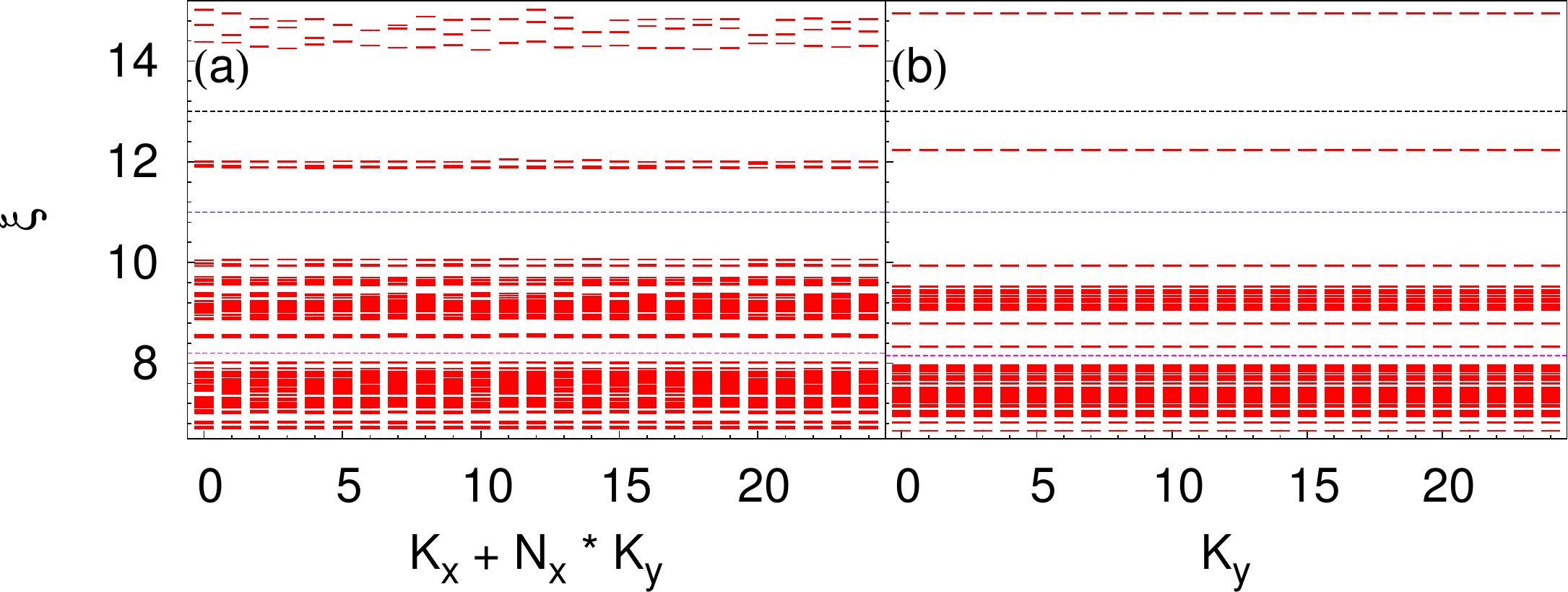}
 \caption{PES of fermionic FCI and FQHE systems with $\nu=\frac{2}{5}$ and $N_A=3$. (a) $10$ fermions on a $5*5$ ruby lattice. (b) FQHE on torus with $10$ fermions and $N_{\Phi}=25$. We observe several gaps in both spectra (each depicted by a line). The counting below each of these gaps obeys
the FQH-to-FCI mapping. The counting below the largest
(and topmost) gap corresponds to the number of Moore-Read
quasihole states. The number of states below the middle gap
is related to the Gaffnian state.}
\label{fig:PESfermions}
\end{figure}

Similarly to the bosonic case, the PES for fermions at $\nu=2/5$ exhibits a non-trivial structure. Fig~\ref{fig:PESfermions} shows the PES for $N=10$ fermions and $N_A=3$ both for the FCI on the ruby lattice and its FQH counterpart. Once again, we observe similar structures between the two spectra.

\end{document}